\documentclass{solarphysics}
\usepackage[optionalrh]{spr-sola-addons} 
\usepackage{epsfig}                      
\usepackage{graphicx}                     
\usepackage{color}                       
\usepackage{url}                         
                        
\newcommand{\ang}{\AA\ }
\newcommand{\gapprox}{\lower.4ex\hbox{$\;\buildrel >\over{\scriptstyle\sim}\;$}}
\newcommand{\lapprox}{\lower.4ex\hbox{$\;\buildrel <\over{\scriptstyle\sim}\;$}}
\newcommand{\arcsec}{\hbox{$^{\prime\prime}$}}

\newcommand{\etal}{{\it et al.}~}
\newcommand{\eg}{{\it e.g.,}~}
\newcommand{\ie}{{\it i.e.,}~}

\newcommand{\aap}{{\sl Astron. Astrophys.}\ } 
\newcommand{\apj}{{\sl Astrophys. J.}\ } 
\newcommand{\apjl}{{\sl Astrophys. J. Lett.}\ } 
 
\newcommand{\ssr}{{\sl Space Science Rev.}\ } 
\def\sp{{\sl Solar Phys.}\ }

\begin{document}
\begin{article}
\begin{opening}
\title{STEREO/Extreme Ultraviolet Imager (EUVI) Event Catalog 2006\,--\,2012}
\runningtitle{STEREO/EUVI Event Catalog}

\author{{Markus J. Aschwanden$^1$} \sep 
	{Jean-Pierre~W\"ulser$^1$} \sep
	{Nariaki~V.~Nitta$^1$} \sep
	{James~R.~Lemen$^1$} \sep
	{Sam~Freeland$^1$} \sep
	{William~T.~Thompson$^2$}}

\runningauthor{M.J. Aschwanden \etal}

\institute{$^1$Solar and Astrophysics Laboratory,
	Lockheed Martin Advanced Technology Center, 
        Dept. ADBS, Bldg.252, 3251 Hanover St., Palo Alto, CA 94304, USA; 
        (e-mail: \url{aschwanden@lmsal.com});
	$^2$ Adnet Systems Inc., NASA Goddard Space Flight Center, 
	Maryland 20770, USA.}

\date{Received ... ; Revised ... ; Accepted ...}

\begin{abstract}
We generated an event catalog with an automated detection algorithm
based on the entire EUVI image database observed with the two 
Solar Terrestrial Relations Observatory STEREO-A 
and -B spacecraft over the first six years of the mission (2006\,--\,2012). 
The event catalog includes the heliographic positions of some 20\,000
EUV events, transformed from spacecraft coordinates to Earth coordinates, 
and information on associated GOES flare events (down to the level of
GOES A5-class flares). The 304 \ang\ wavelength turns out to be the most
efficient channel for flare detection (79\,\%), while the 171 \ang\ (4\,\%), 
195 \ang\ (10\,\%), and the 284 \ang\ channel (7\,\%) retrieve substantially 
fewer flare events, partially due to the suppressing effect of EUV dimming,
and partially due to the lower cadence in the later
years of the mission. Due to the Sun-circling orbits of STEREO-A and -B, 
a large number of flares have been detected on the farside of the Sun, 
invisible from Earth, or seen as partially occulted events.  The statistical 
size distributions of EUV peak fluxes (with a power-law slope of 
$\alpha_P = 2.5\pm0.2$) and event durations (with a power-law slope of 
$\alpha_T=2.4\pm0.3$) are found to be consistent with the fractal-diffusive 
self-organized criticality model. The EUVI event catalog is available on-line 
\url{http://secchi.lmsal.com/EUVI/euvi__autodetection/euvi__events.txt}
and may serve as a comprehensive tool to identify stereoscopically observed 
flare events for 3D reconstruction and to study occulted flare events. 
\end{abstract}


\end{opening}

\section{	Introduction	 			}

Solar flare catalogs from the current solar cycle are available in hard
X-rays from the Reuven Ramaty High Energy Solar Spectroscopic Imager
(RHESSI) during the years 2002\,--\,2013 

(\url{http://hesperia.gsfc.nasa.gov/hessidata/dbase/hessi__flare__list.txt}), 
in soft X-rays from the {\sl Hinode} mission during the years 
2006\,--\,2011 

\url{http://st4a.stelab.nagoya-u/ac.jp/hinode__flare/};
Watanabe, Masuda, and Segawa~2012), and in EUV from the 
Solar Terrestrial Relations Observatory (STEREO) mission for the 
beginning of the mission from December 2006 to July 2008 

\url{http://secchi.lmsal.com/EUVI/euvi__events.txt}; 
Aschwanden \etal 2009). Flare events observed with the Solar
Dynamics Observatory (SDO) are searchable in the web-based browser tool
{\sf iSolSearch} 

\url{http://www.lmsal.com/hek/hek__isolsearch.html},
a window into the Heliophysics Events Knowledgebase (HEK):
Hurlburt \etal2012). An automated flare detection algorithm that
identifies flares in Atmospheric Imaging Assembly (AIA) data 
down to the GOES C-class level is 
operated in real-time as part of the HEK event identification scheme
(Martens \etal2012). The most complete flare catalog for mid-size
to large flares (on the visible hemisphere) is available from the 
Geostationary Operational
Environmental Satellite (GOES), operated by the National 
Oceanographic and Atmospheric Administration (NOAA) and searchable
on the Virtual Solar Observatory (VSO) website 

\url{http://vso.nso.edu/cgi/catalogui/}. 

An automated solar-flare
detection algorithm with about five times higher sensitivity than
the NOAA flare catalog, based on analysis of the 1-8 \ang\ GOES time profiles,
has revealed over 300\,000 solar flare events over the last 37 years of 
1975\,--\,2011 
 
\url{http://www.lmsal.com/~aschwand/GOES/}: Aschwanden and Freeland 2012).

The RHESSI orbit has a duty cycle of about 50\,\% due to its near-Earth
orbit and thus has a corresponding fraction of flare event completeness.
SDO has a duty cycle of almost 100\,\% from its geosynchronous orbit.
{\sl Hinode} has a field of view that does not cover the full Sun, and thus
has a flare event coverage of 51.4\,\% for the X-Ray Telescope (XRT),
24.5\,\% for the Solar Optical Telescope (SOT), and 14.9\,\% for the
EUV Imaging Spectrometer (EIS) (Watanabe, Masuda, and Segawa~2012).

The STEREO mission consists of two twin spacecraft A and B, which circle
the Sun in opposite directions over a period of about 16 years, and thus
have each a field-of-view that rotates the hemispheric view from the
Earth-directed front disk to the Sun's back-side over a time interval
of eight years. Thus, the field-of-view is initially (2006) identical with an
observer's position on Earth, becomes non-overlapping for A and B but
comprises together the full-Sun during the quadrature phase (2010), 
and coincides for A and B at the farside after eight years (2014). 
Consequently, STEREO/Extreme Ultraviolet Imager (EUVI)
has the unique capability to capture a large fraction of flares on the
back-side of the Sun, which are not observable from Earth, or are observed
only partially above some occultation height. Moreover, a large number
of flares are simultaneously observed with STEREO-A and -B, and thus can
be triangulated to quantify their 3D geometric structure. In this article 
we document an automated flare detection algorithm that analyzes the
entire STEREO/EUVI A and B database during the first six years of the
mission (2006\,--\,2012) and provides some statistical results. The full 
EUVI flare catalog is electronically accessible at 

(\url{http://secchi.lmsal.com/EUVI/euvi__autodetection/euvi__events.txt}). 

The EUVI flare catalog information may be particularly useful to identify 
stereoscopically observed flare events (for 3D reconstruction studies),
and to study occulted flare events (regarding their vertical structure).

\section{ 	The STEREO/EUVI Instruments		}

The Extreme Ultra-Violet Imager (EUVI) 
is a normal incidence EUV telescope (Ritchey--Chr\'etien
type) with a 2048$\times$2048 pixel detector, a pixel size of 1.59\arcsec,
a field-of-view out to 1.7 solar radii, and observes in four spectral channels
(Fe {\sc IX} 171 \ang, Fe {\sc XII} 195 \ang, Fe {\sc XV} 284 \ang, 
He {\sc II} 304 \ang ) that span
the $T=0.05\,--\,20$ MK temperature range.  Typical exposure times (at the beginning
of the mission) are 2\,--\,4 seconds for 171 \ang , 4\,--\,16 sseconds 
for 195 \ang , 
and 16\,--\,32 seconds 
for 284 \ang .  The stabilization of EUVI images is accomplished with a
fine pointing system that compensates for spacecraft jitter down to 
the subarcsecond level. Further technical details on the optics,
filters, response functions, and instrument calibration are given by W\"ulser
\etal(2004) and in Section 2 of Howard \etal(2008). 

EUVI images are compressed onboard the spacecraft with the ICER algorithm,
which allows one to trade higher image cadence {\it versus} lower image quality.
Typical compression factors are in the order of 20\,--\,40, without
degrading the image quality more than the photon noise in areas that record
more than a few photons per pixel. The EUVI is typically operated in one of two
modes: the synoptic mode, or the campaign mode. The basic EUVI synoptic mode
has a fast cadence of 2.5 minutes for 171 \ang , a slower cadence of 
10 minutes for 195
and 304 \ang, and 20 minutes for 284 \ang , at the beginning of the mission. 
A higher cadence of 75 s for 171 \ang\ and 5 minutes in the other wavelengths 
was used during special campaigns (for two weeks in May 2007). 
In the later years of the mission, the 171 \ang\ and 284 \ang\ channel were 
operated with low cadences in the order of 120 minutes, which is insufficient 
to detect small flares.

Also, a ring buffer records
EUVI images with a higher cadence during up to four hours per day, which
are downloaded when extra telemetry is available. The telemetry drops stepwise
with increasing distance of the STEREO spacecraft from Earth, reaching a 
minimum at the conjunction point behind the Sun (expected to be in 2014).
The decreasing telemetry also imposes a decreasing cadence of EUV images,
which affects the completeness of flare detection with progress of the mission.

\section{ 	The STEREO/EUVI Event Catalog 		}

\subsection{	Automated Event Detection Method	}

There is a fast and a slow method to automatically detect flare events
in the EUVI database. The quick method reads only the FITS headers of each
image file, which contains the information of the flux maximum (FITS
descriptor {\sf DATAMAX}) in each image, 
from which full-Sun flux--time profiles 
can be created that are sufficient to identify the largest flares 
with some certainty (at GOES M- and X-class level), but without spatial
information. However, for smaller flares (of GOES C-class 
and B-class), the variability of the EUV emission from the quiet Sun, bright
points, explosive events, filaments, and prominences are of comparable
magnitude as small flares, and thus small to intermediate flare events cannot
be identified unambiguously from such full-Sun EUV time profiles. Therefore,
in order to create a deep survey of EUV flare events, we have to choose the
slow method, which means reading and calibrating all observed EUVI images
(which are stored as compressed files and need to be uncompressed for
calibration with the IDL procedure {\sf SECCHI$\_$PREP}). The calibration
with {\sf SECCHI$\_$PREP} converts the observed datanumber per pixel to
a normalized exposure time of 1 second (in units of DN s${-1}$), subtracts a 
dark-current bias, normalizes to open-filter position, and applies a
calibration factor. 

The first step of our automated event-detection algorithm is the generation
of $64\times 64$ daily time profiles extracted from the $2048\times 2048$
pixel images, which represent the average flux $[f_{ij}^{\lambda,sc}(t)]$ 
in macropixels 
with a size of 2048/64 = 32 pixels ($\approx 50\arcsec \approx 36$ Mm),
for each image position $i=1,...,64$ and $j=1,...,64$, wavelength
$\lambda=171, 195, 284, 304$ \ang , and spacecraft $sc=A, B$. 
From the $N=64 \times 64 =4096$ time profiles per wavelength and spacecraft 
we retain only the $N_{\rm loc}=10$ daily time profiles with the highest fluxes, 
which are generally sufficient to cover all active regions with flaring activity,
while the other time profiles mostly belong to the quiet Sun and coronal holes
and are ignored in the further analysis. The Sun has a radius of about 600
pixels in an EUVI image, and thus the ten macropixels with the largest
flux variability cover about a fraction of $q_{\rm Sun}=(10 \times 32^2)/(\pi\ 600^2)
\approx 1\,\%$ of the full-Sun disk, but reduce the amount of time profiles
by a factor of $4096/10 \approx 400$. The accuracy of the heliographic
position inferred from these macropixels (with a size of 32 pixels) is thus
$\pm16$ pixels ($\approx 25\arcsec \approx 18$ Mm $\approx 0.025$ R$_{\odot}$),
or $\pm 1.5^\circ$ heliographic degrees near disk center. 

The second step of the flare detection algorithm consists of
identifying impulsive increases of EUV emission in the previously 
sampled time profiles $[f(t)=f_{ij}^{\lambda,sc}(t)]$. For each
macropixel we detect the maximum flux during a daily time interval
and select for this day those ten time profiles that exhibit the 
largest fluxes (inside the solar disk) among all $64 \times 64 = 
4096$ timeprofiles extracted from each macropixel. Then we add up
these ten time profiles into a synthesized profile.
This combined EUV time profile includes only the brightest
active region areas and has a much higher contrast between impulsive
flux enhancements and the slowly-varying background flux than 
the full-Sun EUV time profiles.

\begin{figure}
\centerline{\includegraphics[width=0.5\textwidth]{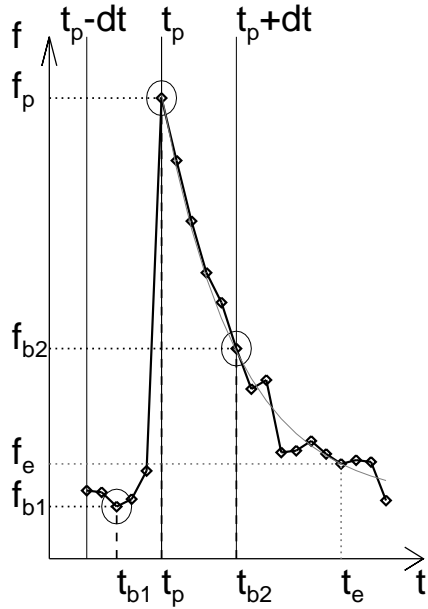}}
\caption{Definition of temporal parameters $[(t_{b1}, t_{b2}, t_p, t_p-dt,
t_p+dt, t_e)]$ and flux parameters $[f_{b1}, f_{b2}, f_p, f_e]$ of an EUV 
event detected in the flux profile $[f^{\lambda,sc}_{x,y}(t)]$.
A maximum time window of $t_p \pm dt$ with half width $[dt=0.5]$ hr
is used for finding the preflare flux minimum $[f_{b1}]$ and the
postflare flux minimum $[f_{b2}]$.}
\end{figure}

For each temporal flux maximum $[f_p=f(t=t_p)]$ at time $t_p$ we determine 
a preflare background flux level $[f_{b1}]$ at time $t_{b1}$ in a time interval
$[(t_p-dt) < t_{b1} < t_p]$, and a postflare flux minimum $[f_{b2}]$
at time $t_{b2}$  with $[t_p < t_{b2} < (t_p+dt)]$ and $[dt]=0.5$ hours 
(Figure 1).  Performing a linear background interpolation 
between the flux minima [$f_{b1}]$ and $[f_{b2}]$,
\begin{equation}
	f_b(t_p)=f_{b1}+(f_{b2}-f_{b1}) {(t_p - t_{b1}) \over (t_{b2}-t_{b1})} \ , 
\end{equation}
we obtain a background level $[f_b(t=t_p)]$ at the peak time $[t_p]$. 
A flare event (at peak time $[t_p=f(t=t_p)]$) is defined by a flux enhancement
that exceeds a background threshold level of $[q_{thresh} = 3\,\%]$,
\begin{equation}
	q = \left( {f_p(t=t_p) \over f_b(t=t_p)} - 1 \right) \ge q_{thresh} \ .
\end{equation}
The threshold level of $q_{thresh}=3\,\%$ has been chosen to optimize the
detection of events down to about the GOES A5-class flare level. 
This sensitivity level
is far above the photon noise level of EUVI data. For instance, the
typical quiet Sun level for the EUVI wavelength channels produces about
100 photons per pixel and second in the 171 \ang\ wavelength (Howard
\etal 2008), which corresponds to $N \approx 10^6$ photons per macropixel
and second, and has a Poisson noise of $\sqrt{N}/N \approx 10^{-3}$ or
0.1\,\%. The least sensitive channel is 284 \ang\ with a quiet Sun level
of four photons per pixel and second, which corresponds to 
$N \approx 4000 $ photons per macropixel and second, and has a Poisson 
noise of $\sqrt{N}/N \approx 0.015$ or 1.5\,\%. Furthermore, since we
analyze only the brightest time profiles from active regions and 
exclude those from the quiet Sun, the significance level of the detected
events is even higher. The significance level of each detected event
can also be inspected {\it a posteriori}, since we provide the background fluxes
$f_b$ and flux enhancements $[(f_p-f_b)]$ in the EUVI event catalog.

Once an event with a significant temporal flux maximum is located
at time $t_p$ in the synthesized time profile of the macropixel $[i, j]$, 
the location [$i_p, j_p$] of the event is identified (from the 
$N_{\rm loc}=10$ 
possible positions) by that macropixel that has the highest flux 
enhancement $[q=f_{ij}^{\lambda,sc}(t_p)/b_{ij}^{\lambda,sc}(t_p)]$
relative to the background flux $[b_{ij}^{\lambda,sc}]$, for which the 
heliographic longitude and latitude [$l_p, b_p$] can be calculated. 

For the time resolution of the extracted time profiles we set a lower
limit of $[\Delta t \ge 0.1$ hour] (or six minutes), 
which is sufficient for flare
detection, but reduces the amount of reading unnecessary images up to
a factor of five during days with the highest cadence (75 seconds) 
in campaign mode.
There are a few gaps in the EUVI/A and B data (mostly at the beginning 
of the mission). EUVI/A started taking data on 4 December 2006, while
EUVI started at 14 December 2006, and the initial testing phase lasted
up to March 2007, before regular temporal cadences were used for observations.
Such data gaps or periods when the spacecraft was rotated produce
discontinuities in the EUV time profiles, which were flagged by 
visual inspection and eliminated in the the flare detection algorithm.
A list of all data gaps (including data drop-outs, bad data, or 
instrumental anomalies) that were excluded in the automated flare
detection algorithm is available in the file 

(\url{http://secchi.lmsal.com/EUVI/euvi__autodetection/euvi__gaps.txt}). 

\subsection{	Heliographic Event Location 		}

All information that is needed to convert heliographic coordinates
from the STE\-REO/A or B spacecraft coordinate system to an Earth-based
coordinate system can be extracted from the FITS descriptors of the
EUVI/A and B images, which includes the pixel numbers $(i_0, j_0)$ 
of the Sun center in the images ({\sf CRPIX1, CRPIX2}), the pixel sizes
$(\Delta x, \Delta y)$ in arcseconds ({\sf CDELT1, CDELT2}), the solar
radius R$_{\odot}$ in arcseconds ({\sf RSUN}), 
the position angle $P$ (in degrees,
or $p=P (\pi/180)$ in radian) of solar North in the image ({\sf CROTA}), 
and the heliographic longitude $l_0$
and latitude $b_0$ of the subsolar point of the STEREO spacecraft 
in a Stonyhurst coordinate system seen from Earth ({\sf HGLON, HGLAT}).
The solar radius in units of image pixels is then
$r_{\odot}=$R$_{\odot}/\Delta x \approx$R$_{\odot}/\Delta y$.
The $(X_{\rm sc},Y_{\rm sc})$ coordinates of a pixel location (i,j) 
in a EUVI image
are then with respect to Sun center in units of solar radii,
\begin{equation}
	\begin{array}{ll}
	X_{sc} &= (i-i_0)/r_{\odot} \\
	Y_{sc} &= (j-j_0)/r_{\odot} \\
	\end{array} \ ,
\end{equation}
or rotated by the position angle $[p]$ to a cartesian spacecraft coordinate 
system with the {\it y-axis} aligned with solar North,
\begin{equation}
	\begin{array}{ll}
	x_{sc} &= X_{sc} \cos{(p)} - Y_{sc} \sin{(p)} \\
	y_{sc} &= X_{sc} \sin{(p)} + Y_{sc} \cos{(p)} \\
        z_{sc} &= \sqrt{1 - x_{\rm sc}^2 - y_{\rm sc}^2 }
	\end{array} \ ,
\end{equation}
which provides the longitude $l_{\rm sc}$ and latitude $b_{\rm sc}$ 
in a spacecraft Stonyhurst coordinate system,
\begin{equation}
	\begin{array}{ll}
	b_{sc} &=\arcsin{(y_{sc})} 	      \\
	l_{sc} &=\arcsin{(x_{sc}/\cos{[b_{\rm sc}]})} \\
	\end{array} \ .
\end{equation}
The Stonyhurst coordinates $[l,b]$ in an Earth-centered coordinate
system with the {\it y-axis} aligned with solar North is given by the
following coordinate transformation, using the heliographic coordinates 
($l_0, b_0$) of the spacecraft suborbital point (Equation~12 in Thompson 2006),
\begin{equation}
	\begin{array}{ll}
	b = \arcsin{(y_{\rm sc} \cos{b_0} + z_{\rm sc} \sin{b_0})} 
	    = (b_0 + b_{\rm sc}) \\
	l = l_0 + \arctan{\left( x_{sc} [z_{\rm sc} \cos{(b_0)} 
	    - y_{sc} \sin{(b_0)}]^{-1} \right)}
	    \approx (l_0 + l_{sc}) \\
	\end{array} \ .
\end{equation}
We provide the spacecraft suborbital coordinates [$l_0, b_0$], the
spacecraft heliographic coordinates $[l_{sc}, b_{sc}]$ of the flare
locations, and the Earth-based heliographic coordinates $[l,b]$ of
the flare locations in the electronically available STEREO/EUVI
event catalog for easy coordinate transformations (\eg to coalign
with {\sl Hinode} or SDO). Further information on coordinate systems used 
for the STEREO/SECCHI instrument suite has been given in 
Thompson (2006, 2010) and Thompson and Wei (2010).

\subsection{ 	STEREO/EUVI Event Catalog Information 		}

The mission-long STEREO/EUVI event catalog, generated by reading the
entire EUVI database and by application of the previously described
automated flare-event detection algorithm, is available at the website

\url{http://secchi.lmsal.com/EUVI/euvi__autodetection/euvi__events.txt}. 

Here we describe briefly the definitions of the parameters that are
listed in this catalog. An extract of one day (8 June 2007) of the
catalog is listed in Table 1, with a graphical representation shown
in Figure 2. The flare catalog can also be read by the IDL routine
{\sf SECCHI$\_$EUVI$\_$FLARECAT} in the {\sf SolarSoftWare (SSW)}.

\begin{figure}
\centerline{\includegraphics[width=1.0\textwidth]{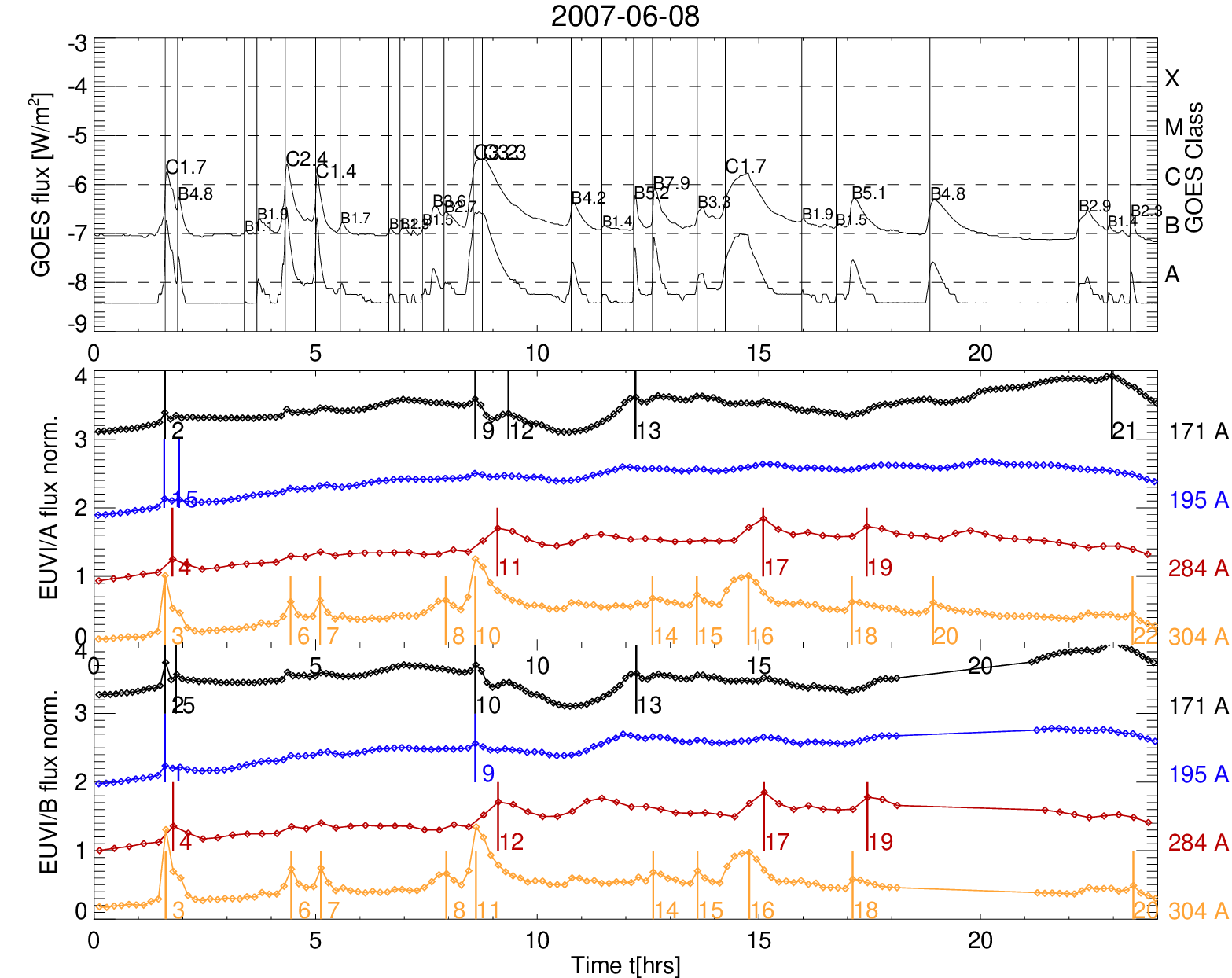}}
\caption{Example of automated EUVI event detection on 8 June 2007,
showing the GOES light curves with automatically detected GOES flares
(top panel), the four EUV time profiles (added from the ten macropixels
with the largest daily fluxes) in 171, 195, 284, and 304 \ang\ with
detected EUVI events from spacecraft STEREO-A (second frame) and
STEREO-B (third frame). Each color corresponds to a different 
wavelength filter. The EUV fluxes are incrementally shifted and 
normalized by $\pm 5$ standard deviations. The parameters 
of each event detected on this day are also listed in Table 1.}
\end{figure}

\begin{table}
\caption{Extract from the STEREO/EUVI catalog 2006\,--\,2012 for the day of 
8 June 2007, containing the date and the peak time $[t_p]$ (hours, minutes), 
the time cadence $[t_c]$ (min), the spacecraft (A,B), a daily event number
per spacecraft ($\#$), the wavelength $[\lambda$ (\ang )], the preflare 
background flux $[f_{b1}]$ [DN $s^{-1}]$, the peak flux $[df=(f_p-f_{b1})]$ 
[DN s$^{-1}$], 
the background change $[db=(f_{b2}-f_{b1}]$ [DN s$^{-1}]$, 
the fluence $[F]$ [$10^3$ DN], 
the rise time $[t_r=(t_p-t_{b1}]$ (minutes), 
the decay time $[t_d=(t_{b2}-t_p)]$ 
(minutes), the GOES peak time difference $[t_g=(t_{GOES}-t_p)]$ 
(minutes), the GOES 
class (G), the spacecraft suborbital longitude $[l_0]$ and latitude $[b_0]$, 
the spacecraft heliographic coordinates of the event 
$[l_{\rm sc}, b_{\rm sc}]$, 
the Earth-based heliographic coordinates $[l,b]$ of the event, and the 
distance from Sun center seen from Earth [$r$], see description in 
Section 3.3).}
\footnotesize
\tabcolsep1pt
\begin{tabular}{lrrrrrrrrrrrrrrrrrrrrrrr}
\hline
Date & $t_p$ & $t_c$ & sc & $\#$ & $\lambda$ & $f_{b1}$ & $df$ & $db$ & F & $t_r$ & $t_d$ & $t_g$ & G & $l_0$ & $b_0$ & $l_{sc}$ & $b_{sc}$ & l & b & r\\
\hline
2007   6   8&   1  35&   7&A&   1& 195&   364.2&   33.0&   26.9&      10&  29&  10&   0& C1.7&    7&   1&   -7&  -6&    0&  -5& 0.179\\
2007   6   8&   1  36&   4&A&   2& 171&   685.3&   49.8&   26.2&      24&  29&   7&   0& C1.7&    7&   1&   -7& -12&    0& -11& 0.250\\
2007   6   8&   1  36&   9&A&   3& 304&  1545.0&  630.7&  226.0&     358&  20&  19&   0& C1.7&    7&   1&   -7&  -6&    0&  -5& 0.179\\
2007   6   8&   1  36&   7&B&   1& 195&   337.1&   22.7&   17.7&       8&  20&  10&   0& C1.7&   -4&   0&    3&  -5&    0&  -6& 0.113\\
2007   6   8&   1  36&   7&B&   2& 171&   579.7&   61.3&   19.7&      38&  22&  29&   0& C1.7&   -4&   0&    3&  -5&    0&  -6& 0.113\\
2007   6   8&   1  37&  10&B&   3& 304&  1479.0&  566.8&  184.8&     327&  20&  20&   0& C1.7&   -4&   0&    3&  -5&    0&  -6& 0.113\\
2007   6   8&   1  46&  19&A&   4& 284&   154.4&   14.8&    8.5&      12&  20&  19&   6& B4.8&    7&   1&   -7&  -6&    0&  -5& 0.179\\
2007   6   8&   1  47&  20&B&   4& 284&   139.4&   11.7&    6.6&      10&  20&  20&   5& B4.8&   -4&   0&    3&  -5&    0&  -6& 0.113\\
2007   6   8&   1  51&   7&B&   5& 171&   586.0&   27.0&   13.3&      41&  22&  22&   1& B4.8&   -4&   0&    3&  -5&    0&  -6& 0.113\\
2007   6   8&   1  55&   9&A&   5& 195&   372.0&   22.3&   14.5&      37&  29&  30&  -2& B4.8&    7&   1&   -7&  -6&    0&  -5& 0.179\\
2007   6   8&   4  26&   7&A&   6& 304&  1658.3&  234.2&   67.6&     167&  19&  20&  -7& C2.4&    7&   1&   -1& -11&    6& -10& 0.206\\
2007   6   8&   4  27&   9&B&   6& 304&  1538.0&  196.9&   52.0&     156&  19&  20&  -8& C2.4&   -4&   0&   11& -10&    6& -11& 0.268\\
2007   6   8&   5   6&   6&A&   7& 304&  1725.9&  179.2&  -16.3&     130&  19&  19&  -6& C1.4&    7&   1&   -1& -11&    6& -10& 0.206\\
2007   6   8&   5   7&  10&B&   7& 304&  1590.0&  152.8&  -25.3&     140&  20&  20&  -7& C1.4&   -4&   0&   11& -10&    6& -11& 0.268\\
2007   6   8&   7  56&   9&A&   8& 304&  1774.4&  135.0&   32.5&     197&  29&  19&  -2& B2.7&    7&   1&   -1& -11&    6& -10& 0.206\\
2007   6   8&   7  57&   9&B&   8& 304&  1635.0&   63.5&  -26.2&     101&  19&  20&  -3& B2.7&   -4&   0&   51& -11&   47& -12& 0.793\\
2007   6   8&   8  36&   5&A&   9& 171&   761.2&   22.1&  -50.5&      32&  15&  22&  -2& C3.2&    7&   1&   -1& -11&    6& -10& 0.206\\
2007   6   8&   8  36&  10&A&  10& 304&  1806.9&  548.6&  206.4&     617&  20&  29&  -2& C3.2&    7&   1&    4& -11&   12&  -9& 0.208\\
2007   6   8&   8  36&  10&B&   9& 195&   393.8&   11.2&   -1.7&      13&  30&  20&  -3& C3.2&   -4&   0&   11& -10&    6& -11& 0.268\\
2007   6   8&   8  36&   6&B&  10& 171&   617.7&   17.6&  -35.3&      33&  15&  22&  -3& C3.2&   -4&   0&   11& -10&    6& -11& 0.268\\
2007   6   8&   8  37&   8&B&  11& 304&  1608.8&  458.3&  235.7&     246&  20&  20&  -3& C3.2&   -4&   0&   16&  -3&   12&  -3& 0.287\\
2007   6   8&   9   6&  20&A&  11& 284&   189.7&   13.7&   10.4&      10&  20&  19& -20& C3.3&    7&   1&    4& -11&   12&  -9& 0.208\\
2007   6   8&   9   7&  20&B&  12& 284&   159.2&    9.6&    7.7&       6&  20&  20& -21& C3.3&   -4&   0&   11& -10&    6& -11& 0.268\\
2007   6   8&   9  21&   6&A&  12& 171&   710.7&   21.3&   -7.1&      37&  22&  22&  99& Z0.0&    7&   1&    4& -11&   12&  -9& 0.208\\
2007   6   8&  12  13&   4&A&  13& 171&   729.8&   59.4&   43.3&      52&  30&   7&  -2& B5.2&    7&   1&    0& -17&    7& -16& 0.306\\
2007   6   8&  12  14&   6&B&  13& 171&   586.8&   28.4&   14.4&      28&  22&  15&  -3& B5.2&   -4&   0&   11& -10&    6& -11& 0.268\\
2007   6   8&  12  36&   9&A&  14& 304&  1837.7&   93.3&   46.4&      98&  29&  30&   0& B7.9&    7&   1&   -1& -11&    6& -10& 0.206\\
2007   6   8&  12  37&   7&B&  14& 304&  1640.4&   69.3&   21.1&      54&   9&  20&  -1& B7.9&   -4&   0&   11& -10&    6& -11& 0.268\\
2007   6   8&  13  36&   7&A&  15& 304&  1854.3&  113.1&    3.4&      85&   9&  29&   0& B3.3&    7&   1&    4& -11&   12&  -9& 0.208\\
2007   6   8&  13  37&   7&B&  15& 304&  1622.6&   97.4&   11.6&      55&  10&  20&   0& B3.3&   -4&   0&   17&  -9&   13& -10& 0.345\\
2007   6   8&  14  46&  10&A&  16& 304&  2014.3&  160.0&  -19.8&     324&  30&  20&  99& Z0.0&    7&   1&    4& -11&   12&  -9& 0.208\\
2007   6   8&  14  47&  10&B&  16& 304&  1836.1&   28.4& -109.8&     114&  20&  20&  99& Z0.0&   -4&   0&   17&  -9&   13& -10& 0.345\\
2007   6   8&  15   6&  20&A&  17& 284&   204.3&    9.6&   -2.2&      12&  20&  19&  99& Z0.0&    7&   1&    4& -11&   12&  -9& 0.208\\
2007   6   8&  15   7&  20&B&  17& 284&   167.8&    8.0&   -0.4&       9&  20&  19&  99& Z0.0&   -4&   0&   17&  -9&   13& -10& 0.345\\
2007   6   8&  17   6&   5&A&  18& 304&  1800.1&   89.7&   61.0&      26&   9&  20&  -1& B5.1&    7&   1&    4& -11&   12&  -9& 0.208\\
2007   6   8&  17   7&   6&B&  18& 304&  1579.8&   74.5&   47.5&      22&   9&  20&  -2& B5.1&   -4&   0&   17&  -9&   13& -10& 0.345\\
2007   6   8&  17  26&  19&A&  19& 284&   194.5&   10.8&    8.4&       7&  19&  20& -21& B5.1&    7&   1&    4& -11&   12&  -9& 0.208\\
2007   6   8&  17  27&  19&B&  19& 284&   163.4&    8.9&    7.2&       6&  19&  20& -22& B5.1&   -4&   0&   17&  -9&   13& -10& 0.345\\
2007   6   8&  18  56&   7&A&  20& 304&  1766.4&  118.2&   56.6&      56&  19&  20&  -4& B4.8&    7&   1&   10& -10&   18&  -8& 0.254\\
2007   6   8&  22  58&   6&A&  21& 171&   848.0&   16.3&  -22.4&      41&  22&  30&  -6& B1.4&    7&   1&    0& -17&    7& -16& 0.306\\
2007   6   8&  23  26&   8&A&  22& 304&  1728.5&   36.5&  -93.8&      94&  20&  30&  -3& B2.3&    7&   1&   10& -10&   18&  -8& 0.254\\
2007   6   8&  23  27&  10&B&  20& 304&  1561.1&   42.7&  -58.5&      69&  20&  30&  -4& B2.3&   -4&   0&   17&  -9&   13& -10& 0.345\\
\hline
\end{tabular}
\end{table}

The Table 1 includes the year, month, and day of observations 
(columns 1\,--\,3), 
the flare EUV peak time $[t_p]$ is given in hours and minutes 
(columns (4\,--\,5), 
the cadence $[t_c]$ during the event, which
is different for each wavelength and spacecraft, is given in units
of minutes (column 6), the spacecraft $(A)$ or $(B)$ where the
flare event is detected (column 7), a successive
event numeration per day and spacecraft (column 8), 
the wavelengths 171, 195, 284, or 304 \ang\ (column 9),
the flare background flux $[f_{b1}]$ in units of [DN s$^{-1}$] as defined in
Figure 2 (column 10), the flare-related flux enhancement $[df=(f_p-f_{b1})]$ 
above the background (column 11), the background flux change 
$[db=(f_{b2}-f_{b1})]$ during
the event (column 12), the (time-integrated) fluence $[F]$ during the
time interval $[t_{b1}, t_{b2}]$ in units of [$10^3$ DN] (column 13),
the rise time $[t_r=(t_p-t_{b1})]$ (column 14), the decay time [$t_d=
(t_{b2}-t_p)$] (column 15), the time delay of the next GOES flare peak 
[$t_g=(t_p^{\rm GOES}-t_p^{\rm EUVI})]$ in units of minutes, within an interval 
of [$\Delta t = 0.5$ hour] (column 16), and the preflare-background subtracted 
GOES class (column 17).
The GOES class may be smaller than the NOAA classification in the
case of weak flares, if the preflare-background level is elevated. 
In the absence of any near-simultaneous GOES flare, a default
GOES class of Z0.0 and delay of 99 is displayed for easier machine readability.
Further we list the
spacecraft suborbital coordinates [$l_0, b_0$] (columns 18\,--\,19), the flare
location [$l_{\rm sc}, b_{\rm sc}$] in spacecraft Stonyhurst coordinates 
(Equation~5)
(column 20\,--\,21),
the flare location $[l,b]$ in Earth-based Stonyhurst coordinates
(Equation~6) (columns 22\,--\,23), and the distance [$r$] of the flare
location from Sun center in the EUVI image (column 24).  
The STEREO/EUVI event catalog is chronologically ordered, containing
the eight merged subcatalogs of flare events independently detected by 
the two spacecraft (A, B) and by the four wavelength filters
(171, 195, 284, 304 \ang ). Thus, some flare events may have up to eight
near-simultaneous detections in this catalog, where the peak times differ
slightly because of the different cadences in each wavelength filter.

\section{ 	STEREO/EUVI Event Statistics 		}

Statistics of detected EUVI events are given in Table 2, showing the number of
events detected in every year from 2006 to 2012, and sorted by wavelengths
(171, 194, 284, 304 \ang ) and spacecraft (STEREO-A, B). The total number of
detected events with both spacecraft in all four wavelength filters amounts
to some $\approx 20\,000$ events, where 79\,\% are detected in 304 \ang ,
10\,\% in 195 \ang , 7\,\% in 284 \ang , and 4\,\% in 171 \ang . The summation in 
the channels demonstrate that most events were detected in the 304 \ang\ 
channel.
Substantially fewer events are detected in the other three channels,
partially because of the EUV dimming at the onset of a CME which can suppress
the impulsive EUV increase at the onset of a flare, and partially
because of the larger cadences used at various times in the 171, 195, 
and 284 \ang\ channels. Our algorithm requires a minimum cadence of 
$t_c \le 0.5$ hr for reliable detection of impulsive flux increases,
and thus channels with longer cadences are not capable of event detection. 
The summations for each year reflect the variability of the solar cycle, 
which exhibited an extended minimum during the years 2008\,--\,2009, 
and produced about ten times more EUV events during the solar maximum 
years of 2011\,--\,2012. 

\subsection{	Event Detection Rates EUVI {\it versus} GOES}

The number of detected EUVI events are presented as histograms in Figure 3,
for each spacecraft [STEREO-A and -B] and wavelength [171, 195, 284, 304
\ang ] separately. For comparison, the monthly event detection rate
of GOES 1\,--\,8 \ang\ flares is also shown (Figure~3, bottom), based on
the automated detection algorithm used by Aschwanden and Freeland (2012). 
These histograms demonstrate that: i) the number of detected events in EUV
and soft X-rays is almost proportional for the 304 \ang\ channel, ii) both 
the EUV 304 \ang\ and soft X-rays 1\,--\,8 \ang\ detection rate show a 
similar modulation through the solar cycle, and 
iii) the EUV detection rates are similar for both spacecraft STEREO-A
and -B. These detection rates reveal for the first time with large 
statistics that flare detection in EUV wavelengths can be as
efficient as in soft X-rays.
It is often stated that flare-event detection in the EUV is hampered
by the effects of EUV dimming, which partially explains that we 
detect a smaller fraction of events in 171 and 195 \ang\ , but a
fair comparison can only be made with equal cadences. During the
STEREO mission, the 171 and 304 \ang\ channels were operated with
a relatively high cadence during the first years of the mission,
while 195 and 304 \ang\ channel were operated with a relatively
high cadence during the later years of the mission.
The 304 \ang\ channel, which is sensitive to chromospheric 
temperatures, appears to be the most suitable channel
for EUV event detection, even at times when the cadences were
similar in the other channels.  

\begin{table}
\caption{Statistics of detected EUVI events, sampled per year (rows),
spacecraft (A,B) and wavelengths (171, 195, 284, 304 \ang ; columns).}
\begin{tabular}{rrrrrrrrrr}
\hline
Year   &     A &     A &     A &     A &     B &     B &     B &      B & Sum  \\
       &   171 &   195 &   284 &   304 &   171 &   195 &   284 &    304 &      \\
\hline
  2006 &    11 &    14 &     6 &     8 &     2 &     6 &     5 &    21 &    73 \\
  2007 &   148 &   101 &   215 &   589 &   124 &    89 &   190 &   507 &  1963 \\
  2008 &    35 &    43 &   221 &   350 &    19 &    22 &   168 &   238 &  1096 \\
  2009 &    38 &    68 &   172 &   559 &    29 &    58 &   168 &   361 &  1453 \\
  2010 &    70 &   216 &    39 &  1352 &    18 &   147 &    19 &  1077 &  2938 \\
  2011 &   108 &   325 &    52 &  2823 &    81 &   210 &    47 &  2470 &  6116 \\
  2012 &    71 &   387 &    60 &  2880 &    66 &   261 &    55 &  2557 &  6337 \\
\hline
   Sum &   481 &  1154 &   765 &  8561 &   339 &   793 &   652 &  7231 & 19976 \\
       & 2.4\,\% & 5.8\,\% & 3.8\,\% &42.9\,\% & 1.7\,\% & 4.0\,\% & 3.3\,\% &36.2\,\% & 100\,\% \\
\hline
\end{tabular}
\end{table}

\begin{figure}
\centerline{\includegraphics[width=1.\textwidth]{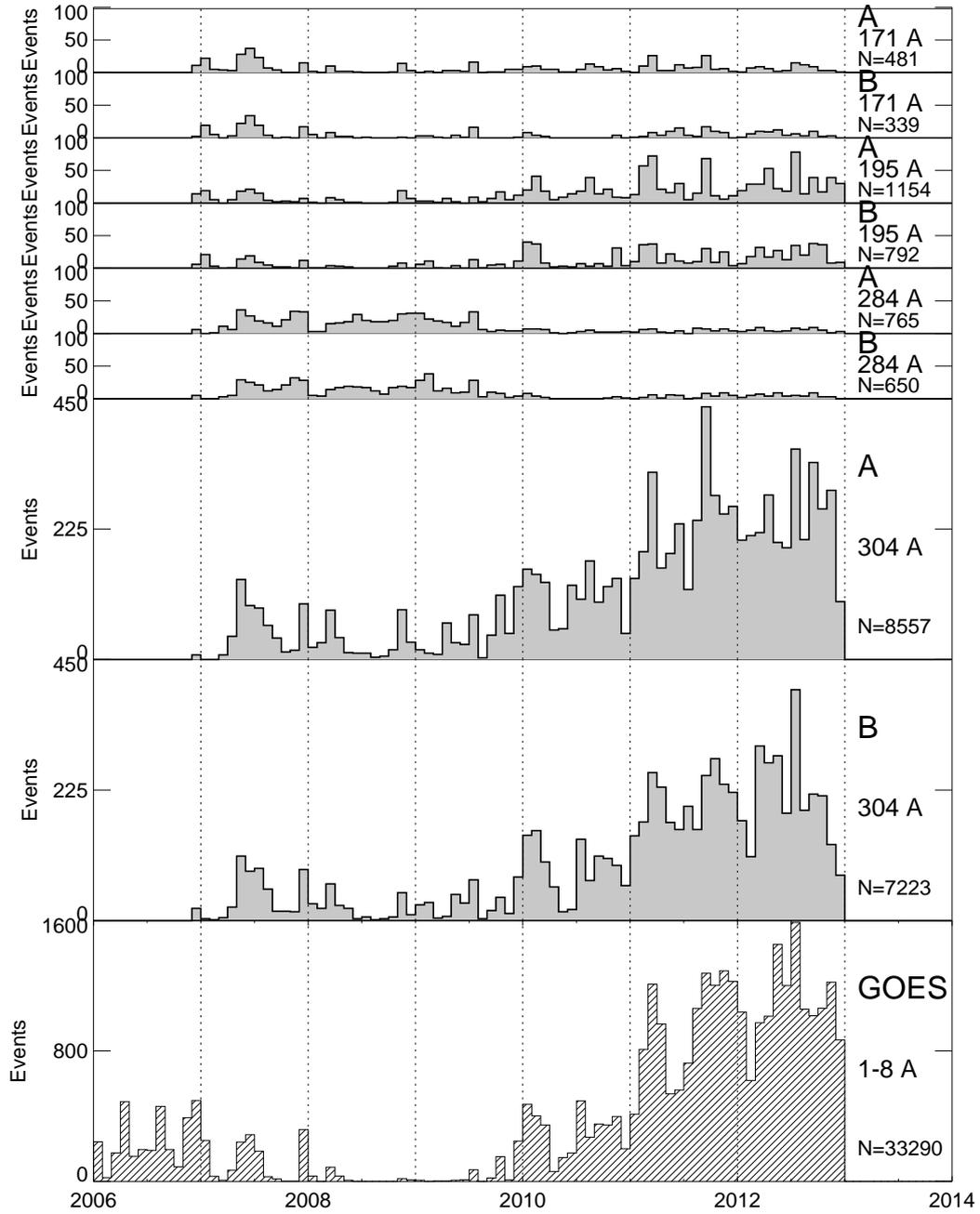}}
\caption{Monthly detection rate of EUVI/A and B events for each
wavelength filter [171, 195, 284, 304 \ang ], and comparison with
the detection rate of GOES events (Aschwanden and Freeland 2012)
in bottom panel.}
\end{figure}

In Figure 4 we explicitly show the near-proportionality of the
(monthly) detection rate of the 304 \ang\ channel of EUVI/A and B
wth the (monthly) detection rate of the GOES 1\,--\,8 \ang\ channel.
The ratio of events detected with GOES 1\,--\,8 \ang\ (with the algorithm of
Aschwanden and Freeland 2012) and those detected in EUVI 304 \ang\
is a factor or $\approx 4.4$ (diagonal line in Figure 4), with a
small scatter for months with high solar activity. During months
of very low flare rates, such as during the extended solar minimum
of 2008\,--\,2009, the relative scatter is larger.

\begin{figure}
\centerline{\includegraphics[width=1.\textwidth]{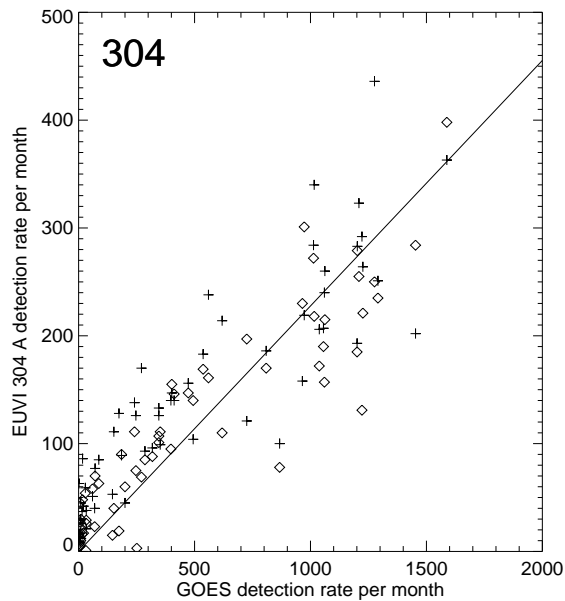}}
\caption{Monthly detection rate of EUVI 304 \ang\
events {\it versus} the GOES 1\,--\,8 \ang\ events. The diagonal line indicates
a proportional detection ratio of 1 EUV event to 4.4 GOES events.}
\end{figure}

\subsection{	Size Distribution of EUV Peak Fluxes 	}

The size distributions of the (automatically detected) background-subtracted 
EUV flux enhancements $[df=(f_p-f_b)]$ are shown in the form 
of log-log histograms
in Figure 5, separately for the spacecraft STEREO-A and -B (left and right
panels) and for the four EUVI wavelengths. Irregardless of the spacecraft
and wavelength, all eight size distributions have a similar power-law slope,
with a mean and standard deviation of $\alpha_P^{\rm obs}=2.5 \pm 0.2$. 
This value 
is similar to a sample of 155 M- and X-class flare events observed with
SDO/AIA in seven (mostly coronal) wavelengths, for which an average 
power-law slope of $\alpha_P = 2.1\pm0.1$ was found (Aschwanden and
Shimizu, 2013). The small 
difference in the slope may be explained by different detection algorithms
and detection criteria (background flux, flux threshold, time interval).  

\begin{figure}
\centerline{\includegraphics[width=1.\textwidth]{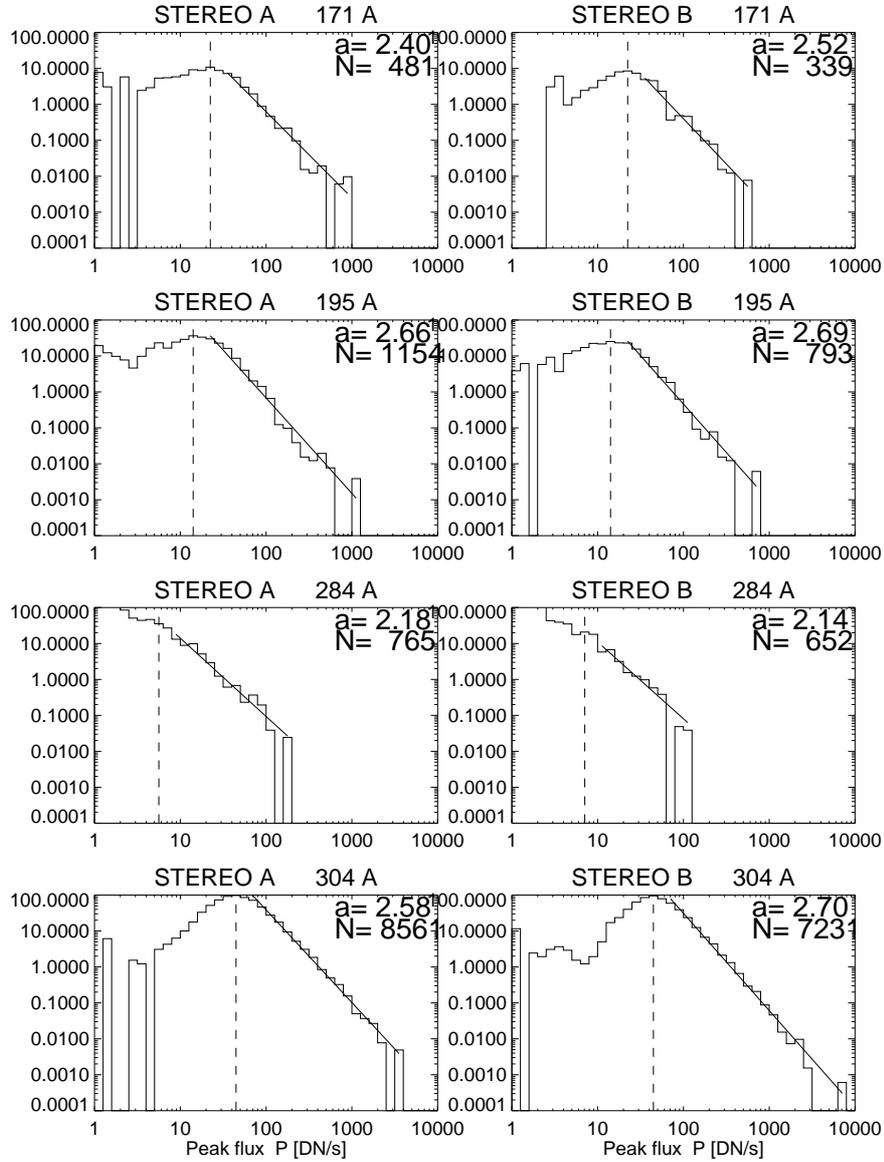}}
\caption{Size distributions of EUV event peak fluxes [$P$], for each
spacecraft STEREO-A and -B (columns) and four wavelength channels (rows).
Power-law fits are indicated with their slopes [$\alpha_P$].}
\end{figure}

\begin{figure}
\centerline{\includegraphics[width=1.\textwidth]{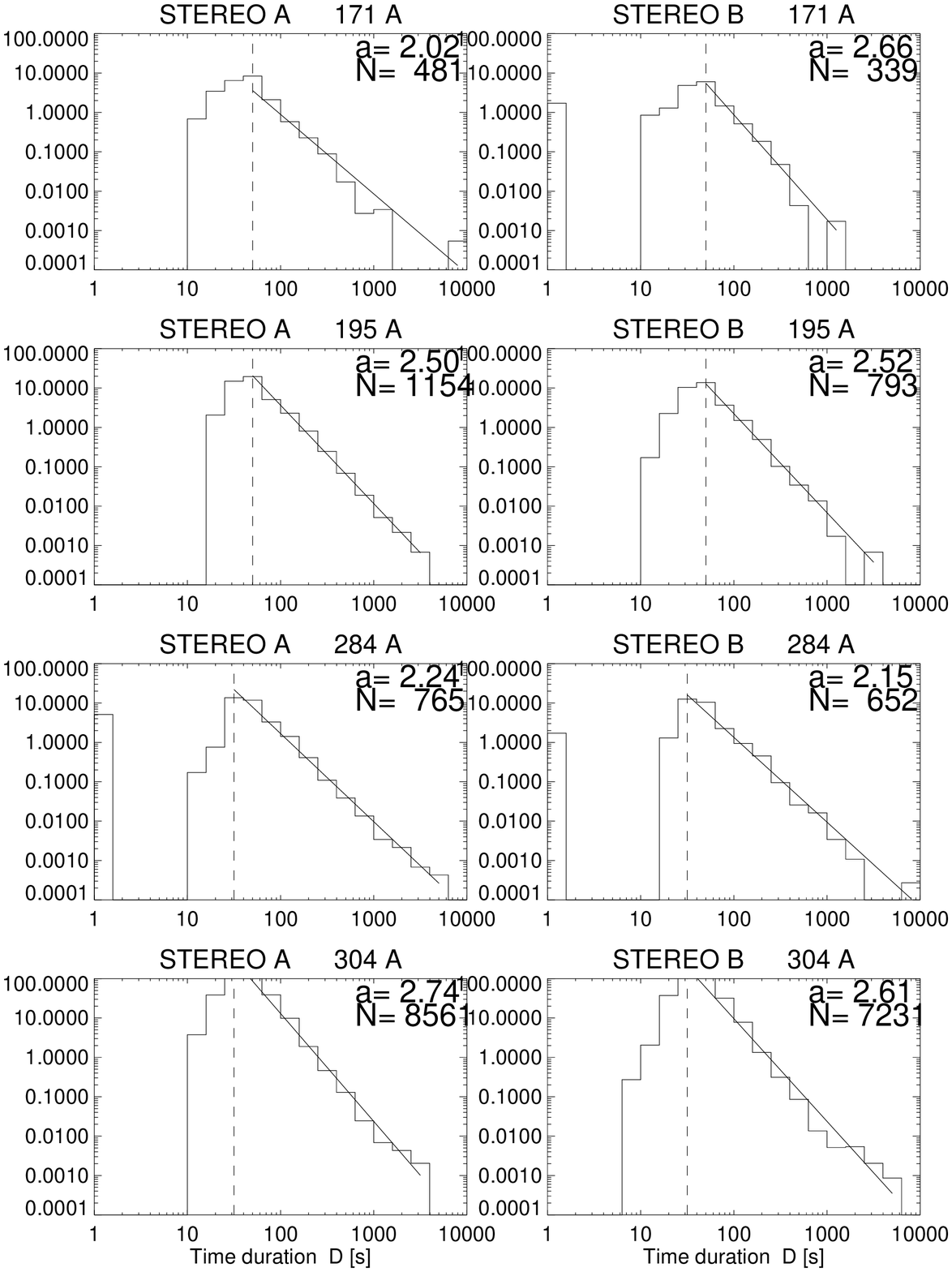}}
\caption{Size distributions of EUV event durations [$\tau$], for each
spacecraft STEREO-A and -B (columns) and four wavelength channels (rows).
Power-law fits are indicated with their slopes [$\alpha_\tau$].}
\end{figure}

\subsection{	Size Distribution of EUV Event Durations }

The temporal parameters we measured for each event consists of the
preflare flux minimum time $[t_{b1}]$, the peak time $[t_p]$, and 
the postflare flux-minimum time $[t_{b2}]$, bound by a maximum time
interval of $[t_p \pm 0.5$ hour].
From these three time points we can define a rise time $[t_r=(t_p-t_{b1})]$,
and a lower limit of the decay time $[t_d=(t_{b2}-t_p)]$, if the background
decreases with time, i.e., $[db=(f_{b2}-f_{b1}) < 0]$. For increasing
background fluxes, however, a more realistic decay time can be
extrapolated by assuming an exponential-decay function and using
the flux [$f_p$] at the peak time [$t_p$] and the flux [$f_{b2}$] 
at [$t_{b2}$[
as constraints to calculate the e-folding decay time [$\tau_e$], and
the end time [$t_e$] when the exponentially decaying flux reaches 
a lower limit of [$q_e=f_e/f_p$], say [$\approx 10\,\%$] of the peak 
flux [$f_p$] (Figure 1). Thus we have the relationships 
\begin{equation}
	{(f_{b2}-f_{b1}) \over (f_p-f_{b1}} = \exp{(-{(t_{b2}-t_p) \over \tau_e})} \ ,
\end{equation}
\begin{equation}
	q_e={(f_e - f_{b1}) \over (f_p-f_{b1})} = \exp{(-{(t_e-t_p) \over \tau_e})} \ ,
\end{equation}
from which the e-folding decay time [$\tau_e$] and the decay time 
[$t_d=(t_e-t_p)$]
can be retrieved. Using this definition of the decay time [$t_d$] we can then
define an event duration from the sum of the rise time and decay time,
\ie [$\tau=t_r+t_d$].
The size distribution of EUV event durations is shown in Figure 6.
The average power-law slope is $\alpha_\tau = 2.3 \pm 0.3$. 

\subsection{Self-Organized Criticality Model}

A theoretical model of a slowly driven fractal-diffusive 
self-organized criticality system (FD-SOC): Aschwanden (2012, 2013)
predicts: i) a power-law length scale distribution [$N(L)$] that is 
reciprocal to the Euclidean volume [$V=L^S$], with Euclidean dimension 
[$S$], postulated by the scale-free probability conjecture,
\begin{equation}
	N(L) \propto L^{-S} \ ,
\end{equation}
ii) a diffusion-type relationship (with diffusion index [$\beta=1$]
for classical diffusion or random walk) between spatial [$(L)$] and
temporal [$T$] scales,
\begin{equation}
	L \propto T^{\beta/2} \ ,
\end{equation}
iii) a fractal geometry for SOC avalanche volumes [$V$] 
\begin{equation}
	V \propto L^{D_S} \ , \quad D_S \approx {(1+S)\over 2} \ ,
\end{equation}
and iv) characterizes the flux--volume relationship between the
observed flux $F$ in a given wavelength and the emitting volume $V$ 
with an additional power-law relationship with index $\gamma$,
\begin{equation}
	F \propto V^\gamma \ .
\end{equation}
Based on these four basic assumptions, the following power-law indices 
[$\alpha_x$] are predicted for the length scales [$L$], volumes [$V$],
time durations [$T$], time-averaged fluxes [$F$], and peak fluxes [$P$],
\begin{equation}
	\alpha_L = S \ , \qquad \qquad \qquad 
\end{equation}
\begin{equation}
	\alpha_V = 1 + (S-1)/D_S \ ,
\end{equation}
\begin{equation}
	\alpha_T = 1 + (S-1) \beta / 2 \ ,
\end{equation}
\begin{equation}
	\alpha_F = 1 + (S-1)/(\gamma D_S) \ ,
\end{equation}
\begin{equation}
	\alpha_P = 1 + (S-1)/(\gamma S) \ .
\end{equation}
Thus, if we set the Euclidean dimension to $S=3$ we expect a mean
fractal (Hausdorff) dimension of $D_3=(1+S)/2=2.0$, a volume 
power-law index of $\alpha_V=2$, a duration power-law index of
$\alpha_T=1+\beta$, and a peak flux power-law index of
$\alpha_P=1+2/(3\gamma)$. In the special case of classical
diffusion we expect $\beta=1$ and $\alpha_T=2.0$. The observed
time duration power-law slope of $\alpha_T=2.4\pm0.3$ is close
to the expected value for classical diffusion.
For comparison, a power-law slope of $\alpha_T=2.02\pm0.04$ was 
found for GOES flares (Aschwanden and Freeland 2013).

If we take the observed peak flux power-law slope of 
$\alpha_P=2.5\pm0.2$ at face value, we infer a flux-volume
power-law index of $\gamma\approx 0.45$. 
In comparison, very similar flux-volume power-law indices 
of $\gamma=0.42\pm0.16$ were found for a data set of 155 M- and X-class
flare analyzed in the 8 AIA wavelengths (Figure 4 in 
Aschwanden and Shimizu 2013). Thus our EUV event detection algorithm
seems to yield flux distributions that are fully consistent with
other statistical studies of EUV events.

\subsection{	STEREO/EUVI vs. GOES Flux Correlations  }

Since we identified a large number of time-coincident events in EUV
and soft X-rays, we can also quantify the flux-flux relationship
between EUV events and soft X-ray events. We find that the
284 \ang\ flux correlates best with the GOES 1\,--\,8 \ang\ flux
(with a cross-correlation coefficient of $ccc \approx 0.5$), which
is sensitive to a temperature range of $\approx 2.0$ MK that is
closest to the GOES 1\,--\,8 \ang\ channel.
The events detected in 171 and 195 \ang\ are probably less correlated 
due to the EUV dimming effects, which reduce the EUV flux at the beginning of
flares during the launch of a CME, and causes a mass loss in the
flaring part of the corona. We show scatterplot of the EUVI 
fluxes with the GOES 1\,--\,8 \ang\ flux in Figure 7. A linear regression
fit with the ordinary least squares bisector method (Isobe \etal1990)
yields a relationship of
\begin{equation}
	f_{EUVI} \propto \left( f_{GOES} \right) ^{0.72\pm0.05} \ .
\end{equation}
In another recent study of flares simultaneously observed with STEREO/EUVI 
and GOES, the 195 \ang\ flux was found to exhibit a good 
correlation with the GOES flux, which was expected due to 
the Fe XXIV line in the 195 \ang\ response function that has a 
sensitivity to hot plasma with a temperature of $\approx 20$ MK,
similar to the response of the GOES 1\,--\,8 \ang\ channel  
(Nitta \etal2013). If we mimic a similar selection of large flares
with a brightness above the GOES C2-class level, we indeed find also a
better correlation of the 195 \ang\ flux with the GOES 1\,--\,8 \ang\ flux
($ccc=0.50$) than the 284 \ang\ flux ($ccc=0.34$). Thus, the overall
correlation behavior between EUV and soft X-ray fluxes depends on 
both the observed wavelength and the flux threshold used in the event
selection, but generally tends to correlate better for large events.

\begin{figure}
\centerline{\includegraphics[width=1.\textwidth]{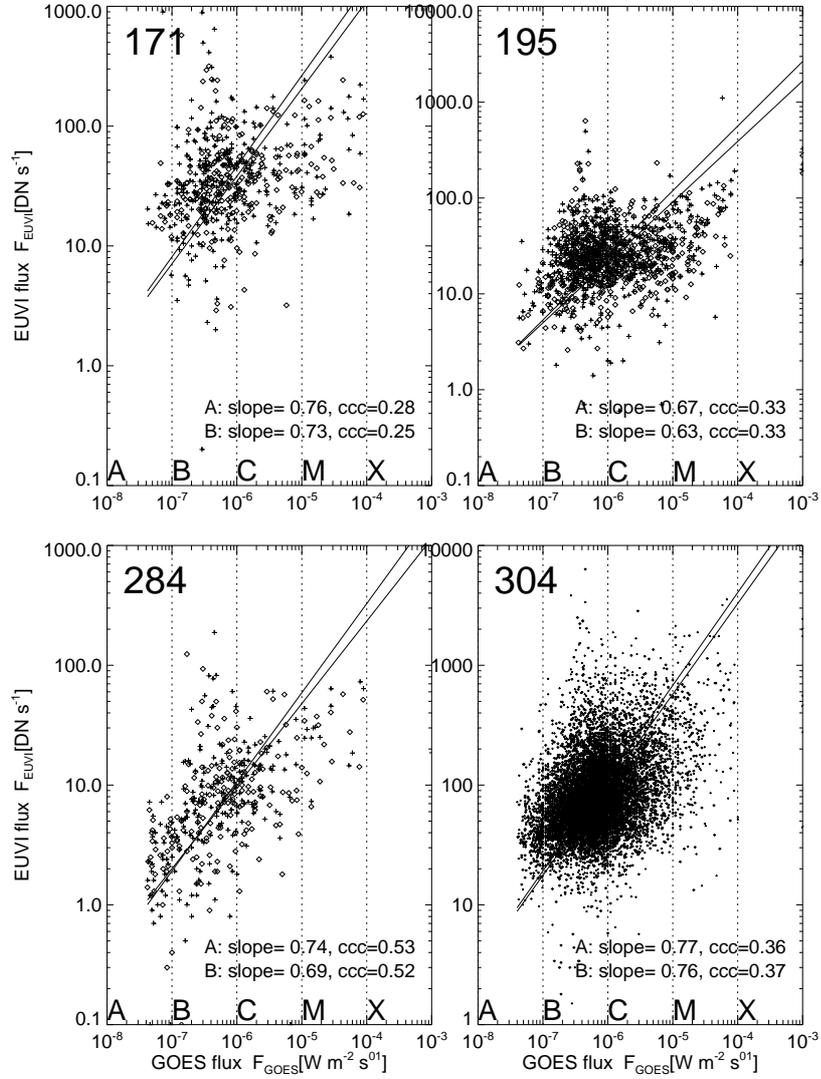}}
\caption{EUVI fluxes [171, 195, 284, 304 \ang ] {\it versus}
the GOES flux of all events simultaneously detected with EUVI and GOES.
The data from spacecraft STEREO-A are indicated with cross symbols,
from STEREO-B with diamond symbols. Separate fits and linear regression
fits for both spacecraft are indicated.}
\end{figure}

\subsection{	Positional Accuracy of Event Locations 		   }

Based on our spatial-localization algorithm that uses macropixels with
a size of 32 image pixels (or $3.0^\circ$ heliographic degrees),
we expect an uncertainty in the heliographic longitude 
[$l \pm 3.0^\circ$] and latitude [$b \pm 3.0^\circ$] at disk center.
The uncertainty increases
for sources that are located near the limb, as seen from one of the two
spacecraft, which can be estimated from the transformation of the
projected longitude to the unprojected longitude. 
If a source is located at a longitude [$l_{\rm sc}$] in the
spacecraft view, an uncertainty of [$dl=3.0^\circ$] projected 
heliographic degrees translates into an error [$\Delta l$] of 
\begin{equation}
	\Delta l \approx {\delta \over \cos{l_{\rm sc}}} \ ,
\end{equation}
Thus for a longitude of [$l_{sc}=45^\circ, 60^\circ$, or $75^\circ$] 
heliographic degrees, we expect an uncertainty of 
$\Delta l  \approx 4.2^\circ$, $6.0^\circ$, or $11.6^\circ$, 
limited by a maximum value of $\Delta_{\rm max}=\arccos(1-\delta)=13.2^\circ$ 
heliographic degrees for positions near the limb, beyond  
[$l_{\rm sc} \gapprox (90^\circ-\Delta_{\rm max}) \approx 76.8^\circ$]. 
Thus, for the chosen macropixel size of [$3.0^\circ$], in our automated
EUV event-detection algorithm we expect a median uncertainty of
[$\Delta l \approx \pm 4.2^\circ$], the value corresponding to the median
longitude of [$l_{\rm sc}=45^\circ$]. This is also comparable 
with the spatial extent of a flare region.

As an experimental test of the accuracy of EUV event positions given
in our EUV event catalog, we compared the listed positions seen from
Earth ($l,b$) with those obtained for a sample of 16 major flares
observed with EUVI behind the limb, compiled in Table 1 of Nitta 
\etal(2013), including events previously analyzed by
Veronig \etal(2010), Dresing \etal(2012), Rouillard \etal(2012),
and Mewaldt \etal(2012). For this sample of 16 major flares we find 
a total of 37 coincident event detections in our EUVI catalog 
(created by an automated detection algorithm),
counting each detection by the two spacecraft and the four wavelength
filters separately. The positional difference between the two flare
sets is found to have a mean and standard deviation of
[$\Delta l = 7.8^\circ \pm 4.0^\circ$], which corresponds approximately
to our theoretical estimate with a median value of
[$\Delta l \approx 4.2^\circ$], given the fact that there is an additional
uncertainty of similar magnitude in the definition of the flare
position, which is given by the spatial extent of an active region,
of the order of a few heliographic degrees. 
  
\section{ 	Summary and Conclusions 			}

We created an automated event detection algorithm that measures
the peak times, EUV fluxes, and heliographic positions of
impulsive EUV events from the STEREO-A and -B spacecraft,
and in each of the four EUVI wavelength channels (171, 195, 284,
304 \ang ) separately. The EUV events in this STEREO/EUVI catalog 
correspond mostly to solar flares detected also in soft X-rays
with the GOES 1\,--\,8 \ang\ channel, down to a level of about 
GOES class A5. A fraction of EUV events without GOES detections may
possibly be associated with filament and prominence activities.
The detection efficiency in the different channels depends strongly
on the temporal cadence used, which necessarily drops with the progress
of the mission due to the increasing distance of the STEREO
spacecraft from Earth and the related reduction of telemetry.
The primary wavelength with high cadence were 171 and 304 \ang\
from 2006 to August 2009, while it was 195 and 304 \ang\ in the later
years of September 2009 to 2012. In addition, EUV dimming
that accompanies the launch of CMEs often prevents the detection
of flares in the 171, 195, and 284 \ang\ channels due to the
partial cancelling of EUV brightening and dimming changes in
EUV flux time profiles. Thus, the 304 \ang\ channel turned out to be
the most efficient channel for EUV event detections (79\,\%),
while the other channels detect a substantially lower number of
EUV events (10\,\% in 195 \ang\ , 7\,\% in 284 \ang\ , and
4\,\% in 171 \ang ); see Table 2.  

The statistics of detected events also strongly varies with the
solar cycle. The lowest number of EUV events (1096 detections)
occurred in the solar minimum year 2007, while the highest number
(6337 detections) was registered in 2012, which is believed
to be close to the maximum of the current solar cycle 24. 
In the overall we detected 19\,976 events, 
which includes multiple detections by the two
spacecraft and four wavelength channels each. If we try to estimate
the number of independent solar EUV events, we can take those from
the most efficient channel and spacecraft, which is 8561 EUV events
detected with EUVI/A in 304 \ang\ (see Table 2). For comparison,
a number of 33,290 independent soft X-ray flare events were
detected with the GOES 1\,--\,8 \ang\ channel with an automated
algorithm that is about five times more sensitive than the official
NOAA solar-flare list (Aschwanden and Freeland 2012). Thus the 
detection ratio of these two automated detection algorithms
is $33,290/8561 \approx 3.9$ (see also Figure 3). The detection
rates of flare events in EUV 304 \ang\ and soft X-ray GOES 1\,--\,8 \ang\
is highly correlated, although the 304 \ang\ emission results from
impulsive brightenings in the chromosphere due to flare-induced
electron precipitation (producing also hard X-rays), while the
soft X-ray emission is mostly believed to be of coronal origin
from post-flare loops that have been filled with chromospherically
heated plasma.

Useful quantitative characterizations of event statistics are
size distributions of the measured parameters. We find for the
occurrence frequency distribution of EUV peak fluxes a power-law
function with a slope of $\alpha_P=2.5\pm0.2$, averaged from all
eight wavelength channels on the two STEREO spacecraft. For the
occurrence frequency distribution of EUV event durations we find
a power-law function with a slope of $\alpha_T=2.3\pm0.3$.
Interpreting the EUV events as nonlinear energy dissipation events
in a slowly-driven fractal-diffusive self-organized criticality
model (Aschwanden 2012, 2013; Bak, Tang, and Wiesenfeld 1987), the measured
power-law indices constrain the spatio--temporal diffusion index
[$\beta = (\alpha_T - 1) = 1.3\pm0.3$] and the flux--volume power-law
index [$\gamma = 2/[3(\alpha_P-1)] \approx 0.45$]. This result is 
close to the case of classical diffusion [$\beta=1$]. The
flux-volume relationship [$F \propto V^{0.45}$] deviates 
significantly from proportionality, as found in another recent
EUV study from AIA (Aschwanden and Shimizu 2013), and thus indicates 
some nonlinear scaling law between the emitted EUV flux and the
flare volume [$V$]. 

The mission-long STEREO/EUVI event catalog of 2006\,--\,2012 is available at:

\url{http://secchi.lmsal.com/EUVI/euvi__autodetection/euvi__events.txt}. 
\ Daily EUVI profiles for both STEREO spacecraft for each of the four
wavelength channels are at

\url{http://secchi.lmsal.com/EUVI/euvi__autodetection/DAILY__EUVI__PROFILES/}, 
synthesized from the most active regions on the solar disk
(combining the ten macropixels with the highest EUV fluxes), which have
a substantially better signal-to-noise ratio than full-disk EUV time
profiles. The EUVI event catalog contains information on the peak times,
rise times, decay times, peak fluxes, background fluxes before and after
the flare, fluences, and heliographic positions of the spacecraft,
the EUV events in spacecraft coordinates, and Earth-based heliographic
coordinates. The heliographic positions allow for an approximate 
localization of flare events within a few heliographic degrees 
(limited by the $3^\circ$ size of the macropixels used in the automated 
detection algorithm). This catalog may be useful for the study of 
steresocopically observed flare events, and occulted flares on the 
farside of the Sun.

\acknowledgements
We thank Fr\'ed\'eric Au\-ch\`ere, Lindsay Glesener, Nat Go\-palswamy, 
Bala Poduval, Andreas Klassen, and David Long for 
helpful discussions and suggestions to create a mission-long STEREO/EUVI
event catalog. Part of the work was supported by the NASA STEREO mission 
under NRL contract N00173-02-C-2035.

\section*{References} 

\def\ref#1{\par\noindent\hangindent1cm {#1}}

\small
\ref{Aschwanden, M.J.: 2012, \aap {\bf 539}, A2.}
\ref{Aschwanden, M.J.: 2013, in {\sl Self-Organized Criticality Systems},
	chapter 13 (Aschwanden M.J., ed.), Open Academic Press, 
	Berlin, Warsaw, p.439, \url{http://www.openacademicpress.de}.}
\ref{Aschwanden, M.J., Freeland, S.L.: 2012, \apj {\bf 754}, 112.}
\ref{Aschwanden, M.J., Shimizu, T.: 2013, \apj, (subm).}
\ref{Aschwanden, M.J., W\"ulser, J.P., Nitta, N., Lemen, J.: 2009,
        \sp {\bf 256}, 3.}
\ref{Bak, P., Tang, C., Wiesenfeld, K.: 1987, Phys.~Rev.~Lett.~{\bf 59}/4. 381.}
\ref{Dresing, N., Gomez-Herrero, R., Klassen, A., Heber, B., Karthavukh, Y., 
	Dr\"oge, W.: 2012, \sp {\bf 281}, 281.}
\ref{Howard, R.A., Moses, J.D., Vourlidas, A., Newark, J.S., Socker, D.G.,
	Plunkett, S.P., and 45 co-authors: 2008, \ssr {\bf 136}, 67.}
\ref{Hurlburt, N., Cheung, M., Schrijver, C., Chang, L., Freeland, S., 
	Green, S.,
	Heck, C., Jaffey, A., Kobashi, A., Schiff, D., Serafin, J., Seguin, R.,
	Slater, G., Somani, A., Timmons, R.: 2012, \sp {\bf 275}, 67.}
\ref{Isobe, T., Feigelson, E.D., Akritas, M.G., Babu G.J.: 1990,
	\apj {\bf 364}, 104.}
\ref{Martens, P.C.H., Attrill, G.D.R., Davey, A.R., Engell, A., Farid, S.,
	Grigis, P.C., Kasper, J., Korreck, K., \etal: 2012, \sp {\bf 275}, 79.}
\ref{Mewaldt, R.A., Cohen, C.M.S., Mason, G.M., von Rosenvinge, T.T., Leske, R.A.,
	Luhmann, J.G., Odstrcil, D., Vourlidas, A.: 2012, In: Solar Wind 13
	Conference Proceedings.}
\ref{Nitta, N.V., Aschwanden, M.J., Boerner, P.F., Freeland, S.L.,
	Lemen, J.R., W\"ulser, J.P.: 2013, \sp (in press).}
\ref{Rouillard, A.P., Sheeley, N.R.~Jr., Tylka, A., Vourlidas, A., Ng, C.K.,
	Rakowski, C., Cohen, C.M.S., Mewaldt, R.A., \etal: 2012, \apj {\bf 752}, 44.}
\ref{Thompson, W.T.: 2006, \aap {\bf 449}, 791.}
\ref{Thompson, W.T.: 2010, \aap {\bf 515}, A59.}
\ref{Thompson, W.T., Wei, K.: 2010, \sp {\bf 261}, 215.}
\ref{Veronig, A.M., Muhr, N., Kienreich, I.W., Temmer, M., Vrsnak, B.: 2010,
	\apjl {\bf 716}, L57.}
\ref{Watanabe, K., Masuda, S., Segawa, T.: 2012, \sp {\bf 279}, 317.}
\ref{W\"ulser, J.P., Lemen, J.R., Tarbell, T.D., Wolfson, C.J., Cannon, J.C., 
	Carpenter,B., and 28 co-authors: 2004, {\it Proc. SPIE} 
	{\bf 5171}, 111.}

\end{article}
\end{document}